\documentclass[]{revtex4}
\usepackage{graphicx}% Include figure files
\usepackage{type1cm} % scalable fonts
\usepackage{lettrine}

\begin{document}

%\draft
\begin{center}
 \textbf{SUBJECT AREAS}\\
Optics and photonics, materials science

\end{center}

\title{Time-resolved dynamics of granular matter by random laser emission}
\author{Viola Folli$^{1,2}$, Neda Ghofraniha$^{3}$, Andrea Puglisi$^{1,2}$, Luca Leuzzi$^{3,1}$ and Claudio Conti$^{1,2}$}
\affiliation{
$^1$Department of Physics, University Sapienza, Piazzale Aldo Moro, 5, 00185, Rome (IT)\\
$^2$ISC-CNR, UOS Roma Sapienza, Piazzale Aldo Moro 5, 00185, Rome (IT)\\
$^3$IPCF-CNR, UOS Roma Kerberos, Univ. Sapienza, Piazzale Aldo Moro 5, 00185, Rome (IT)}
\email{viola.folli@gmail.com}
\date{\today}

\maketitle

 \textbf{Because of the huge commercial importance of granular systems, the second-most used material in industry after water, intersecting the industry in multiple trades, like pharmacy and agriculture, fundamental research on grain-like materials has received an increasing amount of attention in the last decades. In photonics, the applications of granular materials have been only marginally investigated. We report the first phase-diagram of a granular as obtained by laser emission. The dynamics of vertically-oscillated granular in a liquid solution in a three-dimensional container is investigated by employing its random laser emission. The granular motion is function of the frequency and amplitude of the mechanical solicitation, we show how the laser emission allows to distinguish two phases in the granular and analyze its spectral distribution. This constitutes a fundamental step in the field of granulars and gives a clear evidence of the possible control 
on light-matter interaction achievable in grain-like system.}\\

%\pacs{45.70.-n, 05.70.Ln, 42.55.Zz, 81.05.Rm}

%% %%%%%%%%%%%%
\lettrine[lines=3,slope=-4pt,nindent=-4pt]{V}{ertically} shaken granular systems
exhibit a wide ensemble of different phenomenological behaviors. Due
to their peculiar structural conformation, granular materials
\cite{Jaeger1, Jaeger2} are described by a complex phase diagram with
fluid-like \cite{Sano, Melo, Eshuis}, solid-like and liquid-like areas
\cite{Poschel, Poschel2, Paolotti}, depending on the furnished
mechanical energy. Many of the experimental studies on granular
materials uses digital photography, and the analysis is mainly limited
to the external layers of the sample. Various experimental techniques,
able to analyze the internal behavior of a three-dimensional granular
system, have been considered to overcome the opacity of the beads and
give information on the internal arrangements of the grains, as for
example diffusing-wave spectroscopy (DWS) \cite{Pine,Skipetrov,Kim}, magnetic
resonance imaging (MRI) \cite{Ehrichs}, X-Ray microtomography
\cite{Cutress}, and positron emission particle tracking (PEPT)
\cite{Wildman}. These techniques analyze the movement of grains and
give information about their density or related functions.\\In a
previous work \cite{Folli12}, we have shown that a granular material
embedded in a light amplifying medium can emit laser radiation
depending on the configuration of the beads. Here, we employ this
light emission to investigate the granular phases in time resolved
experiments. In particular, concerning with the diffusive-wave spectroscopy method, our experiment gives comparable information with respect to that obtainable with DWS, that is a well-developed and available technique. However, our method can be applied in the studying of temporal coherent properties of granular matter in absorbing media. In fact, in dealing with active and absorbing materials, the phenomenon of absorption dramatically reduces the diffusive-wave signal due to multiply scattered light and nullify the contribution of the longer light paths to the autocorrelation function. Herein, exactly exploiting the absorbing properties of the active medium that surrounds the granular beads, we magnify its response by injecting the sample with a laser source in the resonance region. In this way, the motion of light emitting granulars can be indirectly studied through their scattered laser radiation and its temporal autocorrelation function. So, the usage of laser emission to investigate granular dynamics, presented here, represents a non-invasive and new method that, mutually with DWS, provides a statistical description of the collective dynamical behavior of agglomerations of macroscopic particles in absorbing media.\\At variance with previous work \cite{Folli12}, we consider a completely different system, made by glass beads instead of metallic. The secondary spiked laser emission observed in~\cite{Folli12}, related to the fully reflected nature of metallic beads, is no longer switchable in this sample, but conversely glass beads display a more efficient emission and allow to follow the dynamical properties of the grains. Furthermore, here we analyze a completely new aspect. In the first work, we have observed that the mechanical oscillations of gravity-affected granular sample modify the laser spectra. Here, the temporal variation of laser spectra from oscillated granular is followed dynamically to extrapolate information on the granular phase and we find that the spectral emission is determined by the beads spatial configuration through some non-trivial internal correlations. These are not accessible by means of other techniques and, in some cases, display remarkably oscillating collective modes.\\With respect to standard random lasers (RL)~\cite{Lawandy,Wiersma}, we stress that in "granular" lasers, an agglomeration of macroscopic particles in liquid active media, gravity plays a crucial role in affecting the laser emission spectra and represents a new and fascinating energy source, that jointly with the external mechanical solicitation, allows to tune the laser system above and below its threshold. So, we have created a new kind of laser system, a "granular" laser, that has a tunable emission spectrum. In fact, depending on the externally furnished mechanical energy, an exponentially large number of states of the granular is explored and drives the laser emission.

In our opinion, this work could be
the starting point to use random laser emission as an effective
spectroscopic tool beyond density measurements to investigate internal
collective dynamical modes of shaken and gravity affected
three-dimensional granular systems in liquid solution.\\
%%%%%%%% fig 1 %%%%%%%%%%%%%%%%%%
\begin{figure}[t]
  \includegraphics[width=\columnwidth]{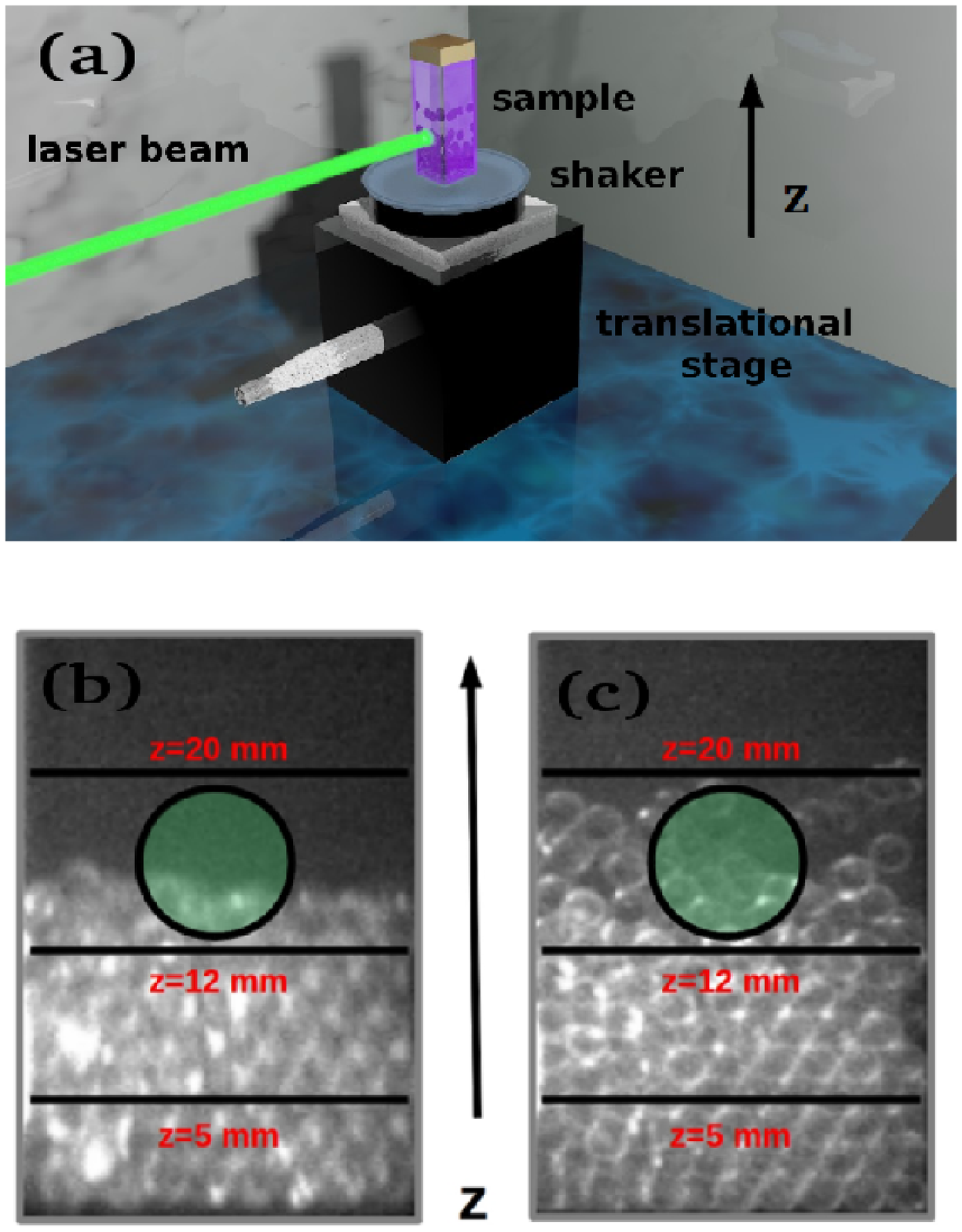}
\caption{(Color online) (a) Sketch of the experiment, (b) grain distribution when the bottom plate oscillates with a displacement $a$ smaller than the bead diameter $D$: $a\cong 0.8 D$; (c) as in (b) with $a\cong 4D$, note the enhanced grain displacements.}
\label{fig1} \end{figure} 
%%%%%%%%%%%%%%%%%%%%%%%%%%%%%%%%%%%%
%%%%%%%%%%%%%%%%%%%%%

\noindent \textbf{Results}\\
\noindent \textbf{Experimental setup.}
In the experiment, the sample consists of 1 mm diameter spherical glass grains dispersed in a liquid Rhodamine B solution. It is placed on a shaker that can vertically vibrate at frequency $f$. The vertical displacement ($z$-direction) $a$ of the vibrating plane is calibrated by an accelerometer. The structure is arranged on a motorized vertical translational stage (maximum travel of $25$ mm). A high energy pump beam is injected on the sample at three different vertical position $z$, fixed when the sample is at rest ($a=0$). A sketch of the experimental setup is shown in Fig.~(\ref{fig1}) panel (a), details are found in Methods. \\
%%%%%%%%% fig 2 %%%%%%%%%%%%%%%%%%
\begin{figure}[h!]
\includegraphics[width=\columnwidth]{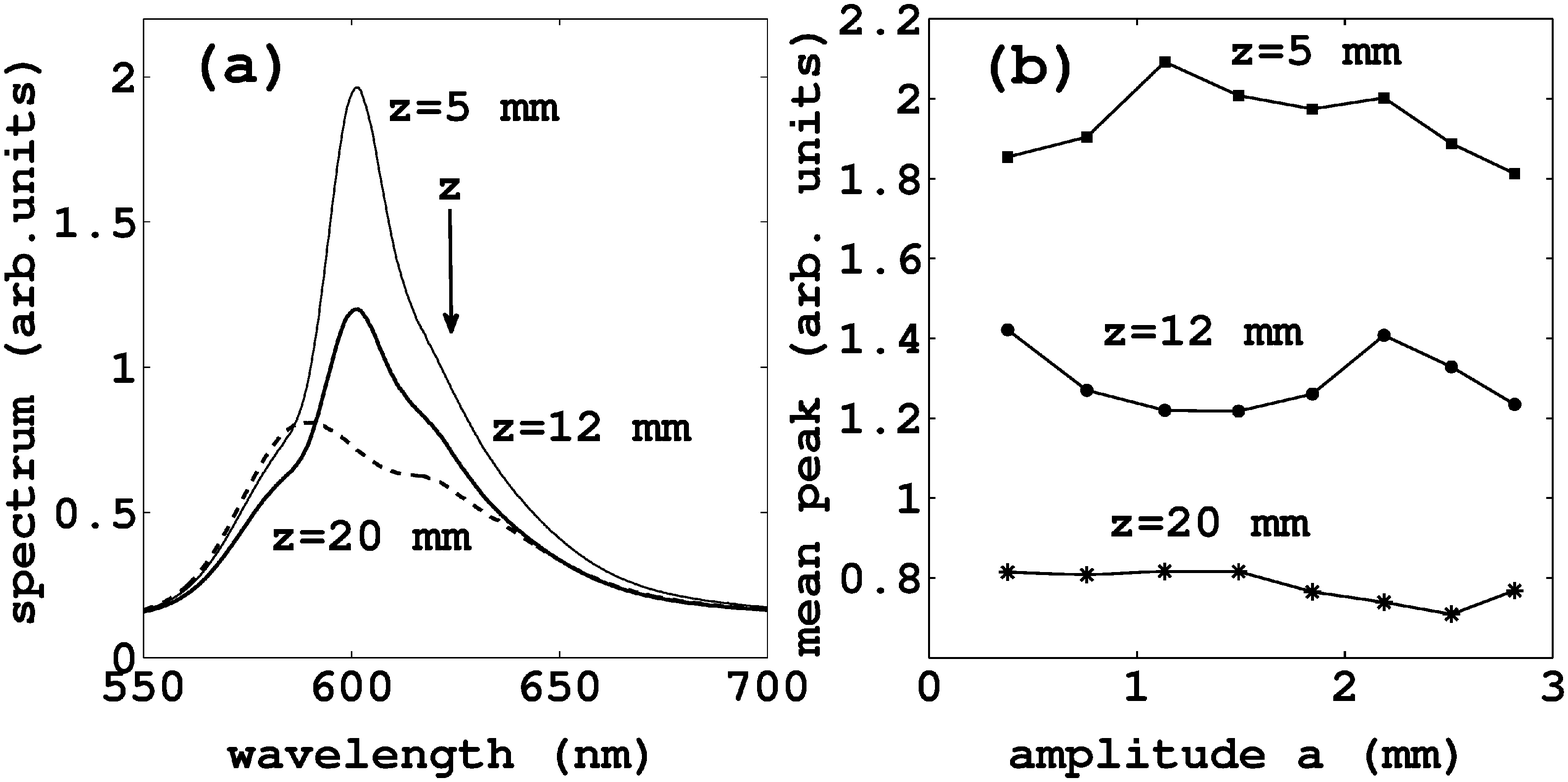}
\caption{(Color online) (a)  Averaged spectra of shaken granular lasers for exposition time of $0.5$ s on 80 accumulations. The input pump energy is $\mathcal{E}=3$ mJ above the random laser threshold, the amplitude and the frequency of the shaking signal are $a=1.49$ mm and $f=70$ Hz ($\Gamma\simeq30$). Continuous thin line gives the laser emission at the bottom layers of granular $z=5$ mm. As the laser spot is moved from the bottom to the top of the container, the laser spectrum becomes less peaked (continuous thick line at $z=12$ mm and dashed line at $z=20$ mm); (b) average spectral peak over $80$ accumulations versus shaker oscillation amplitude for the three different vertical positions, $z=5$ mm (squares), $z=12$ mm (dots) and $z=20$ mm (asterisks).}
\label{fig2} \end{figure} 
%%%%%%%%%%%%%%%%%%%%%%%%%%%%%%%%%%%%%%%%
Varying the vibrational parameters and the vertical position allows to study
the light-grain matter interaction at several structural arrangements of the grains.
Our aim is to use the fast time-scale of emitted light to
investigate the much slower dynamics of such structural arrangements.
 Fig.~\ref{fig1}, panels (b) and (c), show the snapshots of the granular configuration in two
dynamical regimes, with the pump spot size
indicated. Well-known control parameters, namely the normalized shaking acceleration $\Gamma=a\omega^2/g$ (with $\omega=2\pi f$ and $g=9.81$ $m/s^2$), the shaking strength $S=a^2\omega^2/gD$ (with $D$ the bead diameter),
and the layers number are adopted to explore the various phases \cite{Eshuis2, Wassgren, Hsiau, Bizon, Mujica, Mujica2, Sano2}. In this work, we retrieve a phase-diagram by varying the driven frequency $f$ and the shaking acceleration $\Gamma$.\\
%%%%%%%% fig 3 %%%%%%%%%%%%%%%%%%
\begin{figure}[t]
\includegraphics[width=\columnwidth]{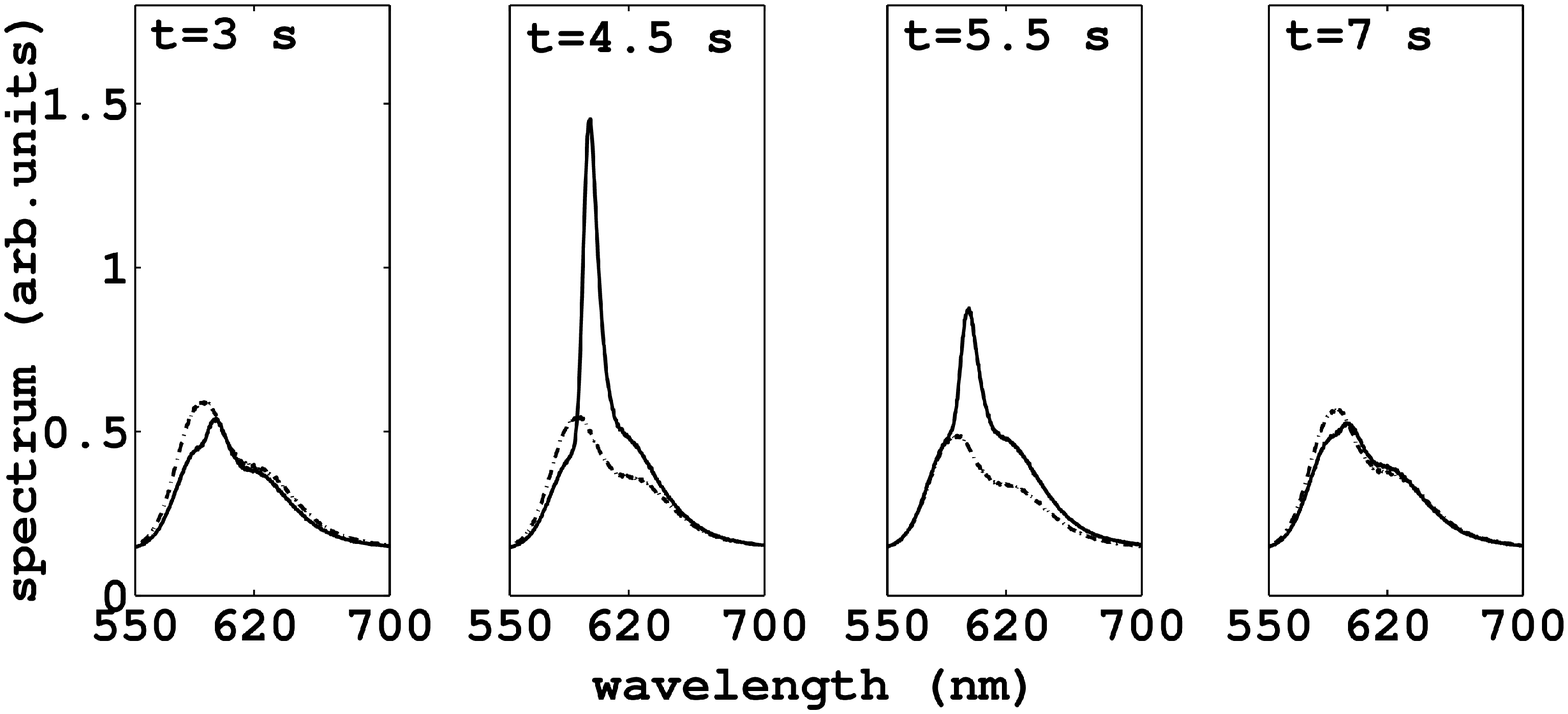}
\caption{(Color online) Five-shots spectra in time, taken at $z=5$ mm (continuous line) and $z=20$ mm (dashed line).\\($\mathcal{E}=3$ mJ, $a=4.96$ mm, $f=70$ Hz, $z=5$ mm).}
\label{fig3} \end{figure} 
%%%%%%%%%%%%%%%%%%%%%%%%%%%%%%%%%%%%

\begin{figure}[t]
\includegraphics[width=\columnwidth]{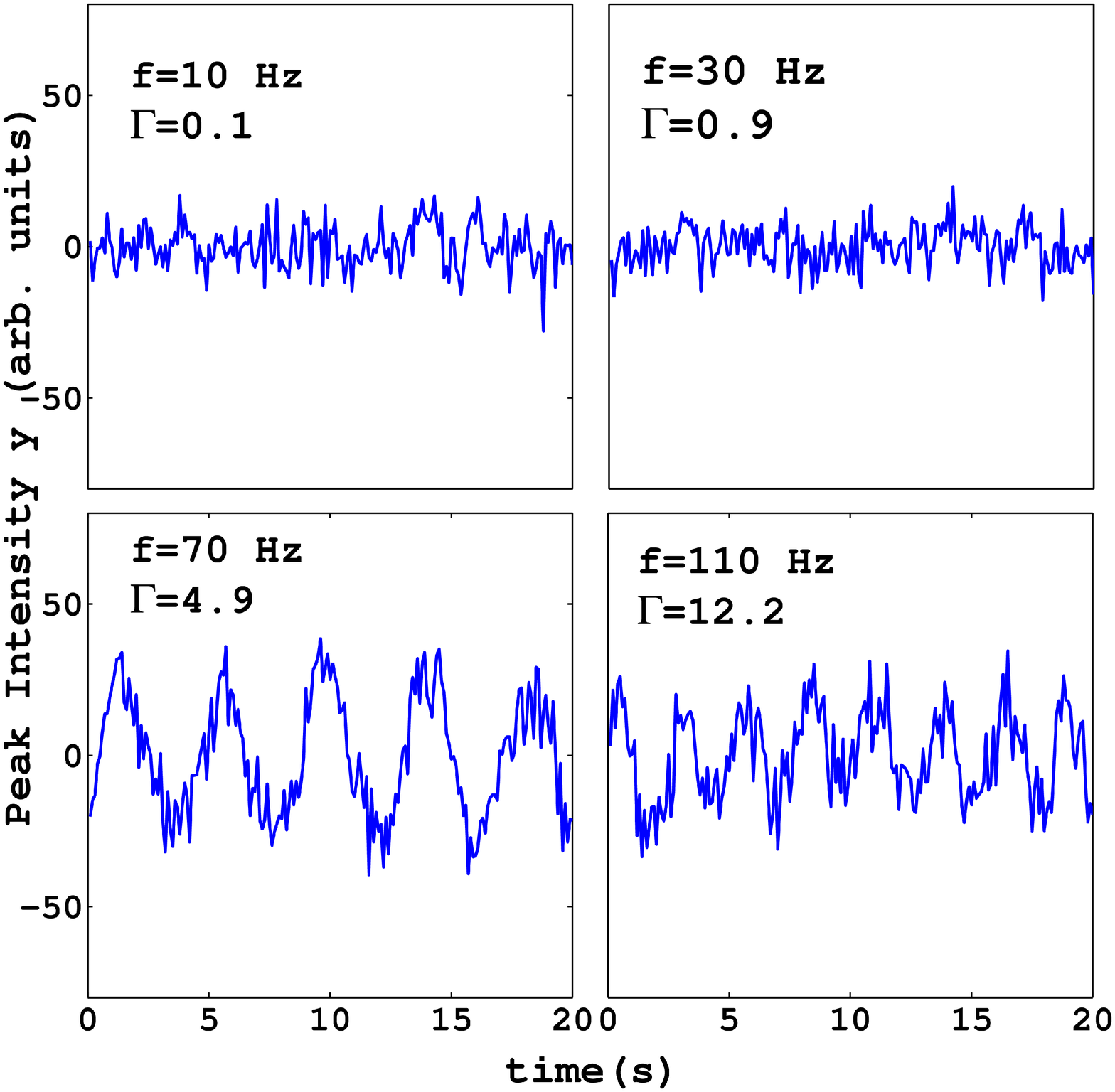}
\caption{(Color online) Peak of one-shot laser emission spectra as a function of time for different frequencies, $a=0.25$. In each plot, we report the corresponding value of $f$ and $\Gamma$.}
\label{fig4} \end{figure}

\textbf{Laser emission.} 
We first determine the vertical position $z$ of the laser beam with respect to the bottom of the cuvette at which the interaction with the light gets optimized, this corresponds to the optimal grain density for the random laser action~\cite{Folli12}. We send a sinusoidal signal with $f=70$ Hz to impress to the granular an oscillation $a\sim 1.5D$, ($\Gamma=30$). In this way a dynamical regime with a relevant grain motion is reached. We fix three different heights $z$, at the bottom ($z=5$ mm), at the edge ($z=12$ mm) and at the top region ($z=20$ mm) of the granular. Note that the vertical position $z$ is taken with respect to the position of the base of the cuvette at rest, the three chosen positions are separated by the pump spot size.\\The averaged spectra in Fig.~\ref{fig2}, panel (a), show the laser is more efficient at $z=5$ mm (continuous thin line). For larger $z$, the laser efficiency is reduced (continuous thick line at $z=12$ mm and dashed line at $z=20$ mm) because the granular density decreases with $z$ and the laser cavity becomes more lossy as $z$ is increased (reduced scattering strength). Figure~\ref{fig2} (b) shows the averaged spectral peak as a function of the shaker displacement; the peak is more pronounced at the bottom region of the cuvette.
As shown in \cite{Folli12}, the laser emission of granular shaken laser depends on the furnished mechanical energy and is determined by the specific instantaneous granular configuration. Here, we investigate the granular collective temporal evolution by following the dynamics of its laser emission spectra. We start addressing the time dependence of the granular spectra in Figure~\ref{fig3}. The reported spectra are averaged over five shots (CCD exposure time $0.5$ s) and are taken after $t=3,4.5,5.5$ and $7$ s from an arbitrary time at two different heights. The spectra are substantially distinct due to a significant variation of the mean grain density. We find that the spectra evolve periodically, and Figure~\ref{fig3} (a), (b), (c) and (d) correspond to an oscillation period. At first glance, the occurrence of this oscillation is  unexpected, indeed the time-scale of the laser emission is much faster than the grain motion, and should be related to the average density that is constant. If the laser emission oscillates, it must sense higher order structural properties. To show evidence that the time dynamics of laser emission allows to discriminate the dynamical evolution of granular matter and it is able to reveal a collective motion underlying the observed density modulation, we proceed as follows: we fix $z=5$ mm, corresponding to maximal spectral variation (Fig.~\ref{fig3}), we choose the pump energy at $\mathcal{E}=3$ mJ above the random laser threshold, we fix the oscillation amplitude at $a=0.25$ mm and vary the frequency $f$. During the vibrations of the grains, we collect $200$ laser spectra with exposition time $0.1$ second (corresponding to a time interval of $36$ seconds time interval because of the readout time $t_{Rout}=0.08$ s of the CCD camera) for each measurement. We analyze the experimental results by plotting the peak of each consecutive spectrum as a function of time. This analysis is repeated by making a frequency $f$ scan from $10$ to $120$ Hz with a step of $10$ Hz (Fig.~\ref{fig4}) in order to explore a wide range of $\Gamma$, namely $0.1\leq\Gamma\leq14.5$.  
%%%%%%%% fig 4 %%%%%%%%%%%%%%%%%%
%%%%%%%%%%%%%%%%%%%%%%%%%%%%%%%%%%%% 
We observe that when the shaker
plate is subject to a shaking acceleration approximately less than $5$, the whole granular barely moves; correspondingly, the
dynamical evolution of the laser spectra is negligible, as the mean
bead density is constant with time. When the peak vibration amplitude $\Gamma$ is increased, the mechanical solicitation on the granular overcomes the
gravitational force, the capillary forces between grains and liquid, and the constraints imposed by the sides of the
cuvette. In such a regime, the granular is fluidized and the laser
spectra reveal an interesting and unreported dynamics. In
Fig.~\ref{fig4}, we show the dynamical evolution of the laser peak
signal $y$ at different frequencies (similar results are obtained when
measuring the spectral width, not reported). The laser spectra
allow to distinguish two distinct phases of the granular: the first
regime, observed for $\Gamma\lesssim 5$ ($f<70$ Hz in Fig.~\ref{fig4}), is a solid
phase (Fig.~\ref{fig1}b) where the beads follow the motion of the
plate but do not change their relative positions. The second regime,
activated when $\Gamma\gtrsim 5$ ($f\geq70$ Hz), is an oscillating-phase
(Fig.~\ref{fig1}c) in which the beads move in the whole volume of the
container and site exchanges between beads take place. In the latter
phase, the pronounced density fluctuations trigger the laser above and
below threshold, periodically. As a result, the spectra alternate
between fluorescence and random laser. We call this oscillating state
a liquid-like phase.
%%%%%%%% fig 5 %%%%%%%%%%%%%%%%%%
\begin{figure}[t]
\includegraphics[width=\columnwidth]{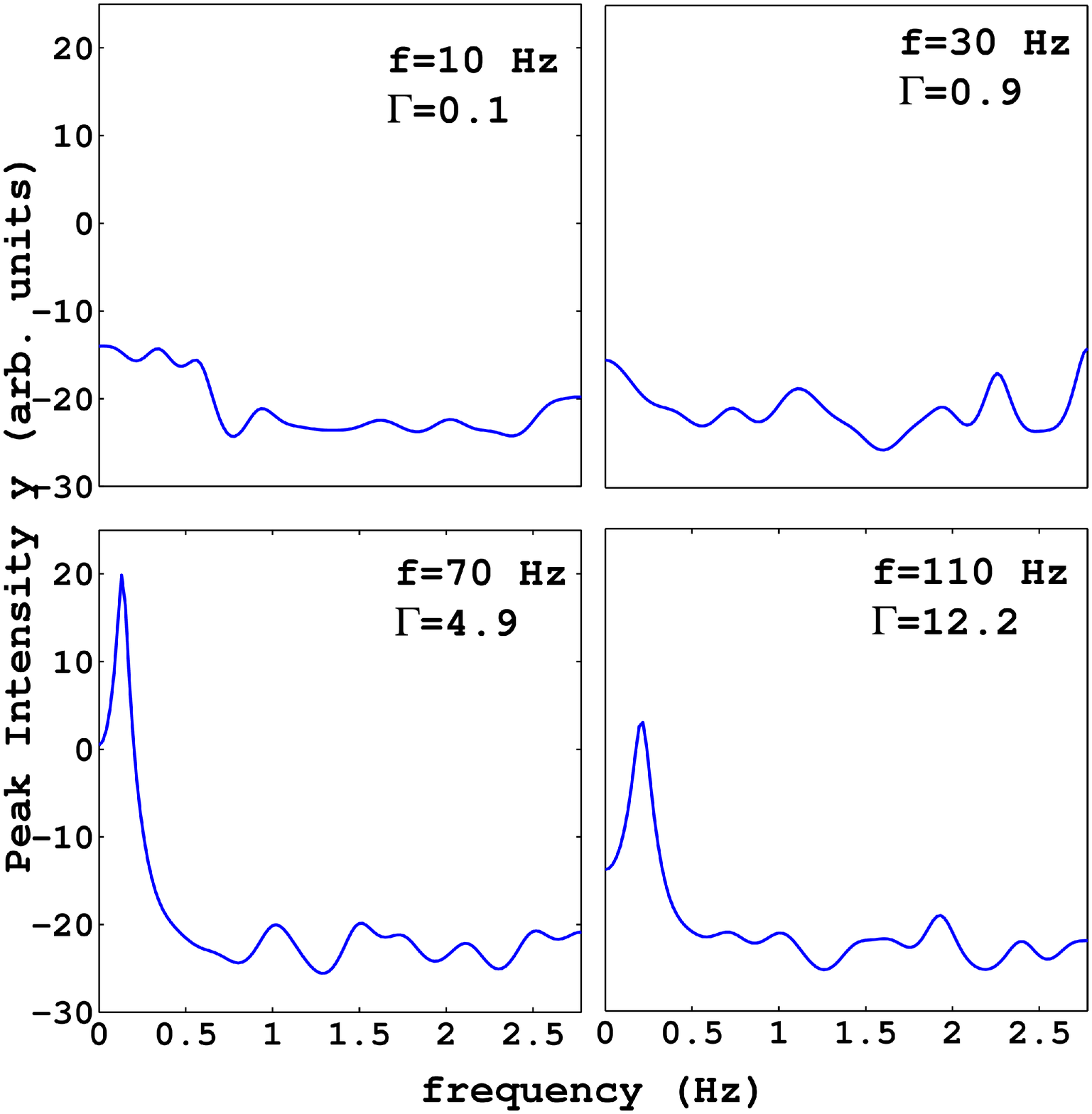}
\caption{(Color online) Fourier-domain analysis of the temporal signals in Figure~\ref{fig4}, with shaking frequency $f$ and $\Gamma$ indicated.}
\label{fig5} \end{figure} 
%%%%%%%%%%%%%%%%%%%%%%%%%%%%%%%%%%%% 
To further investigate this
effect, we analyze the spectral information in the dynamics of the
oscillating response of Fig.~\ref{fig4}. We perform a Fourier-domain
analysis of the temporal signals in Fig.~\ref{fig4}, reported in
Fig.~\ref{fig5}. We then extract the fundamental frequency of
oscillation by fitting the normalized data $\eta=y-\overline{y}$,
where $\overline{y}$ is the mean value of the temporal evolution, by a
single sinusoidal function. The analysis reveals two phases with
well-separate solid-like and liquid-like areas; the phase transition
is evidenced when the amplitude of the fit is plotted as a function of
the frequency (Fig.~\ref{fig6}). There is evidence of a resonant peak
at $f=70$ Hz in the oscillating region. The granular response is hence
not flat but shows an internal structure with preferential mechanical
modes. In the inset, we report the fitted frequency of the oscillating
phase versus the driven frequency. The interpolated frequencies are two orders of magnitude smaller
than known typical granular oscillations frequencies, among which the so-called "bouncing-bed" regime~\cite{Eshuis}. The latter is
an oscillation of the whole granular like a solid, which, in dry
samples, has a characteristic value $f_{bb}$ given by
$f_{bb}=f/2$.\\To exclude subsampling and stroboscopic effects, we
performed acquisition of spectra at different sampling times. We
did not find relevant changes. The fact that the measured frequency is
lower than $f_{bb}$ in dry granulars can be ascribed to the hosting
liquid. The physics of liquid-immersed granulars is in many respects
unknown, the analysis of the dynamic granular laser emission indicates
the presence of very low frequency vibrational modes.

%%%%%%%% fig 6 %%%%%%%%%%%%%%%%%%
\begin{figure}[t]
\includegraphics[width=\columnwidth]{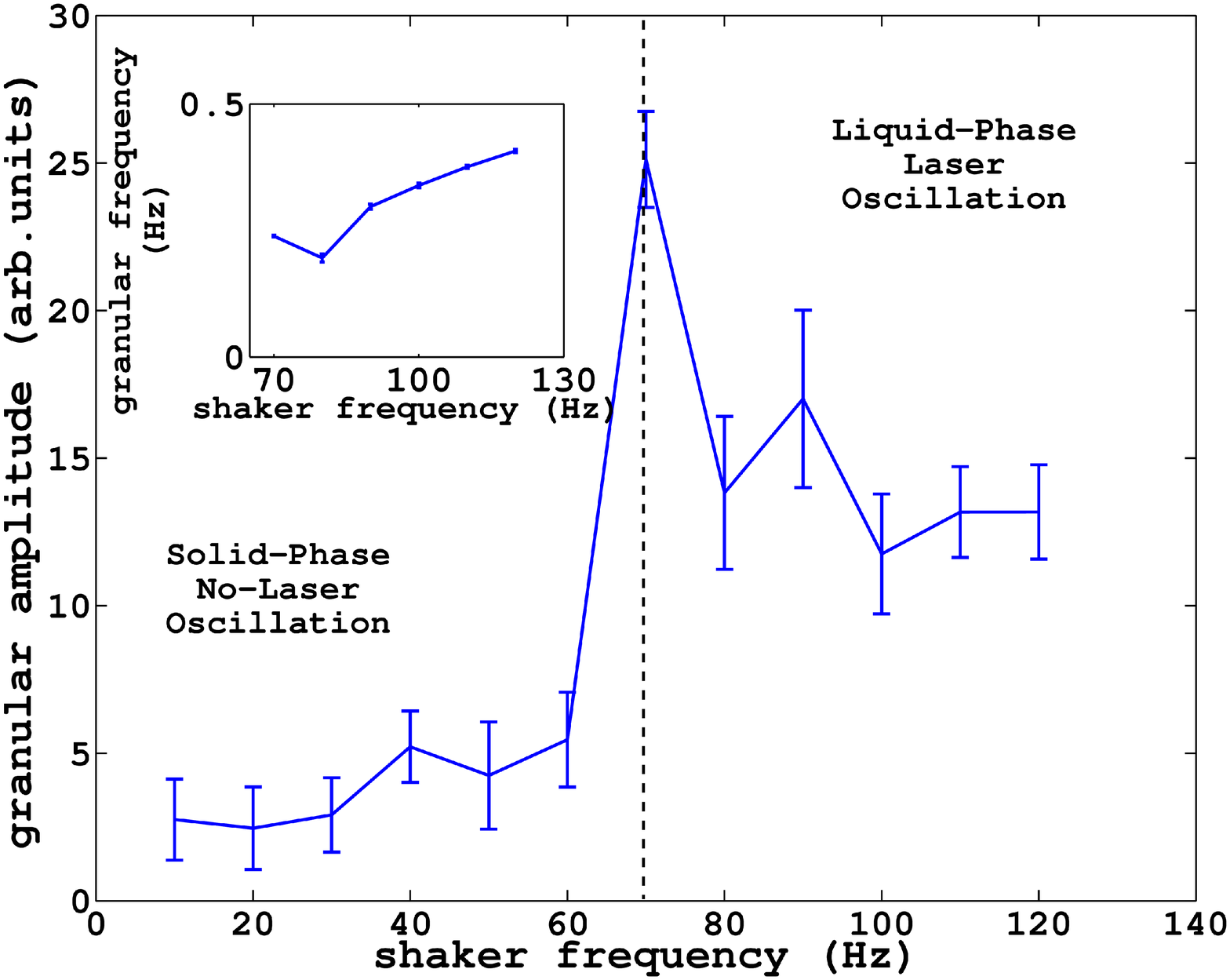}
\caption{(Color online) Fitting amplitude of the normalized signals $\eta$ as a function of the frequency. In the inset, it is reported the frequencies of the dynamical evolution of the laser spectra in the oscillating-phase ($f\gtrsim70$ Hz).}
\label{fig6} \end{figure} 
%%%%%%%%%%%%%%%%%%%%%%%%%%%%%%%%%%%%
%%%%%%%% fig 7 %%%%%%%%%%%%%%%%%%
\begin{figure}[t]
\includegraphics[width=\columnwidth]{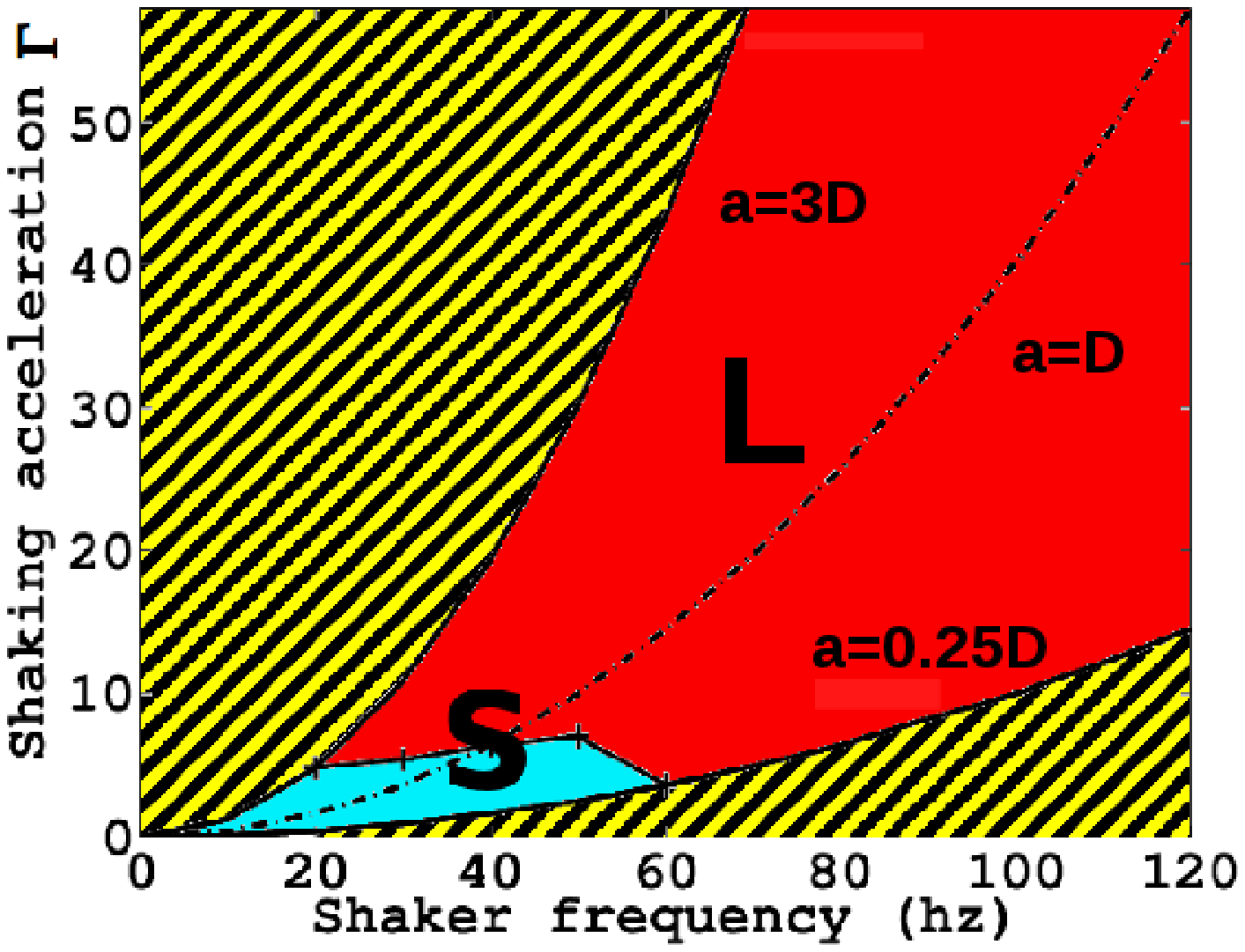}
\caption{(Color online) Phase diagram of glass bead with $D=1.0$ mm with Rhodamine B. The region in red (dark gray) corresponds to the liquid-like (G) phase, related to the laser emission spectra oscillations, the region in blue (light gray) indicates the solid-like (S) phase. The region filled with stripes represents areas out of the range of linearity of shaker response, and hence not explorable.}
\label{fig7} \end{figure} 
%%%%%%%%%%%%%%%%%%%%%%%%%%%%%%%%%%%% 
To better address the liquid-like
state and the origin of the laser oscillations, we perform further
investigations and measure a phase-diagram (see figure~\ref{fig7}) to
characterize the features of our sample. The adopted order parameter
is the peak $\hat{P}$ of the spectral analysis in Fig.~\ref{fig5}. We
identify the liquid-like phase to the regime in which the peak amplitude $\hat{P}$ is about three times greater than the background and correspondingly, the solid-like phase when no relevant peaks are observed. We see that the liquid-like
state (oscillatory dynamics) is present also below the line at which
the plate displacement equals the diameter of the beads ($a\cong D$),
and the oscillations of the laser seem to result from the density
modulation of the granular, and to depend on the grain inertia further
enhanced by the presence of the liquid active medium (dyed methanol)
that slows down the beads.\\ 

\noindent \textbf{Discussion}\\
In \cite{Folli12}, we had shown the configurational dependence of laser emission from the shaken granular. In this work, we follow dynamically this dependence to extract informations about the structural phases of the grains. We have found evidence of a periodic time dependence of laser emission from a granular sample when it is opportunely shaken. This temporal correlation is related in a non-trivial manner to the dynamical phases of the granular. These oscillations are not related to some convective motion of a few glass particles that, absorbing light, are carried periodically through the laser spot. In fact, a setup like that we used, with an aspect ratio of $L/H\simeq 1$ and with $L\simeq 10D$, can not support granular convection rolls or undulations. We stress that a rigorous understanding of the observed
phenomenon is still missing but we believe that the granular laser
oscillations have to be related to a sort of pulsation in the
configurational statistical properties of the out-of-equilibrium
liquid granular system. 
In conclusion, we first provide the
experimental evidence that the laser emission furnishes information
on the dynamics of the granular system. Not only we distinguish two
granular phases, the static and the dynamical regimes, but also
provide evidence that laser analysis encodes the spectral structure of
the granular collective motion. Random lasing hence provides a
quantitative tool to investigate collective motion in granular
matter. 

Developments include applying these results to
study granular systems with more complex phase-diagram and link with
statistical mechanics. Additionally, the rich and collective behavior
of granular matter may allow to implement a new class of random laser,
tunable by the mechanical control.\\ 

\noindent \textbf{Methods}\\
\textbf{Sample.} Our sample consists of about
$N=1.194$ grams (about $1200$ particles) of $D=1$ mm diameter spherical
dielectric spheres. The grains are immersed in a solution of Methanol
and Rhodamine B (concentration $1$ mM, fluorescence peak around 590 nm), and are contained in a square
cuvette of side length $L=10$ mm $= 10D$. The sample height in the
vertical direction is $H\simeq 10D$ in order to have an aspect ratio
$L/H\simeq 1$. We choose this regime to reduce
the variety of supported granular phases which become more complex for $L/H\gg 1$~\cite{Eshuis},
\cite{Eshuis2}.\\ \textbf{Setup.} The container is placed on a shaker that can
vertically vibrate. The applying force controlling the shaker is
sinusoidal at frequency $f$.
The vertical displacement $a$ of the granular cuvette has been
measured by employing a piezoelectric accelerometer. A pump laser beam, a Q-switched 532-nm Nd:YAG laser (10 Hz
repetition rate, 7 ns pulse duration and spot size 6 mm), is used, and
the spot position $z$ with respect to the cuvette bottom is set by a vertical
motorized 25 mm translational stage.
Spectral emission is measured by an electrically cooled CCD
array detector (operating temperature $-70$ degree Celsius).

%\bibliography{MEGAbib}
%\bibliographystyle{nature}
%\end{document}

\vspace{1cm}
\noindent\textbf{Acknowledgments}\\
The research leading to these results has received funding from the
European Unionʼs Seventh Framework Programme FP7/2007–2013 under the
European Research Council ERC grant agreement n.201766 and under the
People Programme (Marie Curie Actions) REA grant agreement No. 290038;
from the Italian Ministry of of Education, University and Research
(MIUR) under the PRIN Projects No. 2009P3K72Z and No. 2010HXAW77-008
and under the Basic Research Investigation Fund (FIRB/2008)
Program/CINECA Grants No. RBFR08M3P4 and No. RBFR08E7VA.\\

\noindent\textbf{Author contributions}\\
C.C., N.G., V.F. designed research, V.F. performed experiments, analyzed data and wrote the manuscript, C.C., N.G., A.P. and L.L. critically reviewed it. All authors gave their approval for the final version of the paper.\\  

\noindent \textbf{Additional information}\\
The authors declare no competing financial interests.

\noindent \textbf{Corresponding author}\\
Correspondence should be addressed to V.F. (email: viola.folli@gmail.com)\\ 

\end{document}